\documentclass[journal]{IEEEtran}
\usepackage{amsthm}
\usepackage{algorithm}
\usepackage{url}
\usepackage{multirow}
\usepackage{amssymb}
\usepackage{subfig}
\usepackage{algorithm}
\usepackage[noend]{algpseudocode}
\usepackage{amsfonts}
\usepackage{filecontents}
\usepackage{mathtools}
\usepackage{xcolor}
\definecolor{rv1}{rgb}{1.0, 0.44, 0.37}
\definecolor{rv2}{rgb}{0.4, 1.0, 0.0}
\definecolor{rv3}{rgb}{0.0, 0.75, 1.0}
\definecolor{rvt}{rgb}{0.75, 0.75, 0.75}
\usepackage{soul}

\newcommand{\etal}{\textit{et al.}}

\newtheoremstyle{exampstyle}
{7pt} % Space above
{7pt} % Space below
{\itshape} % Body font
{} % Indent amount
{\bfseries} % Theorem head font
{.} % Punctuation after theorem head
{.5em} % Space after theorem head
{} % Theorem head spec (can be left empty, meaning `normal')
\theoremstyle{exampstyle}

\algrenewcommand\algorithmicindent{2em}%

\ifCLASSINFOpdf
\usepackage[pdftex]{graphicx}
\graphicspath{{FigureRoot/}}
\else
\fi
\newcommand{\hb}{{\bf h}}

\newcommand{\xb}{{\bf x}}

\newcommand{\yb}{{\bf y}}
\newcommand{\zb}{{\bf z}}

\newcommand{\Dc}{{\cal D}}

\newsavebox\mybox

\begin{document}
	%\title{Designing Fast Gradual Sparse Attacks to \\Deep Neural Networks}
    \title{Network Intrusion Detection with Limited Labeled Data Using Self-supervision}
	
	\author{S. Lotfi, M. Modirrousta, S. Shashaani, and M. Aliyari Shoorehdeli
		%\thanks{This paragraph of the first footnote will contain the date on which you submitted your paper for review. It will also contain support information, including sponsor and financial support acknowledgment. For example, ``This work was supported in part by the U.S. Department of Commerce under Grant BS123456.'' }
		\thanks{S. Lotfi (e-mail: slotfi72@email.kntu.ac.ir) is with the Intelligent Information Processing LAB (IIPLAB), K.N.Toosi University of Technology, Tehran, Iran.
		
		M. Modirrousta (e-mail: mohammadbc@email.kntu.ac.ir) is with the Fault detection and Identification LAB (FDI), K. N. Toosi University of Technology, Tehran, Iran.
		
		S. Shashaani (e-mail: sh.shashaani@email.kntu.ac.ir) is with the Intelligent System LAB (ISLAB), K. N. Toosi University of Technology, Tehran, Iran.
		
		M. Aliyari Shoorehdeli (e-mail: aliyari@email.kntu.ac.ir) is with the Faculty of Electrical Engineering, K. N. Toosi University of Technology, Tehran, Iran.}}
	

	\markboth{Journal of \LaTeX\ Class Files,~Vol.~14, No.~8, August~2015}%
	{Shell \MakeLowercase{\textit{et al.}}: Bare Demo of IEEEtran.cls for IEEE Journals}



	\maketitle
	
	\begin{abstract} 
With the increasing dependency of daily life over computer networks, the importance of these networks security becomes prominent. Different intrusion attacks to networks have been designed and the attackers are working on improving them. Thus the ability to detect intrusion with limited number of labeled data is desirable to provide networks with higher level of security. In this paper we design an intrusion detection system based on a deep neural network. The proposed system is based on self-supervised contrastive learning where a huge amount of unlabeled data can be used to generate informative representation suitable for various downstream tasks with limited number of labeled data. Using different experiments, we have shown that the proposed system presents an accuracy of $94.05\%$ over the UNSW-NB15 dataset, an improvement of $4.22\%$ in comparison to previous method based on self-supervised learning. Our simulations have also shown impressive results when the size of labeled training data is limited. The performance of the resulting Encoder Block trained on UNSW-NB15 dataset has also been tested on other datasets for representation extraction which shows competitive results in downstream tasks.
\end{abstract}
	
	\begin{IEEEkeywords}
		Intrusion, self-supervised, labeled data, Transferability.
	\end{IEEEkeywords}
	\IEEEpeerreviewmaketitle
	\section{Introduction}
	
% \begin{abstract}%
\noindent
% \section{prior_art}
	
Recent technological advances such as cloud computing, big data, Artificial Intelligence (AI), the Internet of Things (IoT), and other technologies have made people's lives increasingly dependent on the networks and the internet.

Industrial Control Systems (ICSs) have demonstrated a trend of interconnection leading to the Industrial Internet, an interconnected system that integrates people, machines and things. However, network attacks continue to increase, putting the security of ICSs that depend on data transmissions at risk. Anderson initially proposed intrusion detection as part of the network security system \cite{anderson1980computer}. He described an intrusion attempt as an unauthorized attempt to access and manipulate information and as a result, the system becomes unreliable.

Intrusion detection technology in industrial control networks or traffic data analysis can promptly predict and take active defensive measures when other systems are compromised. There are two types of Intrusion Detection Systems (IDS), host-based \cite{rbace:icsa2000} and network-based \cite{hamed2018network} systems. A host-based IDS collects information from within a particular computer, such as system calls, application logs, file access logs, file-system modifications, password files and other host activities \cite{bridges2019survey}. However, a network-based intrusion detection system collects raw network packets from different segments of a network and analyzes them systematically for signs of intrusion \cite{yang2019deep}.

From another point of view, IDS can be divided into two categories: (1) signature-based systems and (2) anomaly-based systems \cite{jyothsna2011review}. A signature-based IDS detects destructive code based on predefined patterns called signatures. This method is effective for static detection with a low False Positive Rate (FPR). Manually updating the signature database and the inability to detect unexpected attacks are significant challenges in these systems. Thus attacks with zero-day vulnerabilities are harder to detect \cite{buczak2015survey}. On the other hand, the anomaly-based system detects destructive behavior based on deviations from standard functioning. These systems can detect zero-day attacks. Despite their advantages, anomaly-based systems have the disadvantage of having difficulty detecting normal traffic accurately leading to high FPR. Hybrid systems combine the two above techniques to achieve a high Detection Rate (DR) with low FPR \cite{sethi2020context}.

Classical Machine learning (ML), a versatile tool to map input features $\xb$ to target outputs $\yb$ based on learning over a set of $\{\xb_i,\yb_i\}_{i=1}^N$ pairs, can effectively fit to IDS. Classical ML algorithms, mainly rely on shallow input-output mappings, have been widely used as IDS core systems \cite{dllb15}. Support Vector Machines (SVM) \cite{yerong2014intrusion}, Decision Trees (DT) \cite{sahu2015network}, k Nearest Neighbors (kNN) \cite{aburomman2016novel}, Random Forest (RF) \cite{farnaaz2016random}, Naive Bayes (NB)  Classifier \cite{yang2018modified}, and shallow Artificial Neural Networks (ANN) \cite{saber2017performance}, to mention but a few, are some examples of shallow ML algorithms used in IDS. 

While classical ML algorithms generally lead to easily explainable rules with robust mappings, they lack efficient handling of large datasets and their performance saturates while the dataset size is increasing \cite{dllb15, mantere2012challenges}. The revolution of the digital world and incredible speed of data generation and the bottleneck in classical ML algorithms in handling large datasets lead to the development of deep learning (DL) architectures \cite{dllb15}. DL architectures are a hierarchy of shallow mappings and have shown to be universal approximators \cite{auall20}. In DL models, the classical handcrafted features are replaced with trainable layers, which result in better generalization performance and avoids performance saturation while increasing the dataset size \cite{sarker2021deep}.

DL models generally use deep neural networks (DNNs) architecture \cite{dllb15}. From the architecture point of view, neural networks can be categorized into feed-forward (FFNN) and recurrent (RNN) groups. While FFNNs have no loop in their architectures, RNNs have at least one loop which mainly affects their learning procedure \cite{dlgb16}. The loop in the RNNs makes them suited for inputs coming from a sequential process (e.g.\ speech, text and time series) and the architecture of FFNNs adjust them to model batch signals (e.g.\ Image) \cite{dlgb16}.

While deep FFNNs are simple extension of shallow neural networks by adding layers, their successful performance originate from the introduction of Restricted Boltzmann Machines (RBMs) that can intelligently pre-train layers' weights using unlabeled data \cite{aflho06}. This direction was then followed by the introduction of Autoencoders (AEs) \cite{bengio2006greedy}. Long Short-Term Memory (LSTM) network is a specialized implementation of RNNs. LSTMs maintain the sequence information in a unit named state to be used in the future of the sequence. They have been used in many applications including text translation, image recognition, speech recognition and anomaly detection problems in time-series sequence data \cite{hochreiter1997long}. Both FFNNs and RNNs have been widely used in IDS \cite{javaid2016deep, Yin2017ADL, keshk2017privacy, chandak2019analysis, li2019lstm, zhu2019mobile, erfani2016high, fiore2013network, wang2015applications, ashraf2020novel}.

The reliance of supervised learning algorithms over labeled dataset and its dependency to the target task, limit its application for many input types where huge amounts of unlabeled data is accessible via the digital world. On the other hand RBMs and AEs that use unlabeled data for pre-training are suited to shallow architectures \cite{rlwol18}. 

The aforementioned limitation led to the introduction of self-supervised learning (SSL) algorithms. SSL utilizes deep architectures to generate high level representation suitable for various downstream tasks in a fully unsupervised manner. The resulting representation is then fed to a head architecture to generate the target output for a specific downstream task and trained using a limited number of labeled training pairs \cite{rlwol18}.

Contrastive learning is a widely used SSL framework where the objective is to train a network such that it can identify different views to a phenomenon while contrasting between different phenomena \cite{chen2020simple}. As an example in the machine vision tasks, the architecture is trained so as to identify augmented versions of original input image where the augmentation methods include cropping, rotating, gray-scale transformation, etc. This framework has been applied to intrusion detection \cite{wang2021network}. While the resulting performance is considerable, the authors have not investigated the effect of labeled dataset size over the performance of the proposed system and their justification for the selected augmentation methods is not clear.

In this paper, we investigate the use of contrastive learning framework for intrusion detection. Our main contributions are as follows:
\begin{itemize}
    \item \textit{Data Augmentation:} As a well-justified method for augmentation, we propose to use masking. This selection makes the contrastive learning framework robust against missing elements of input pattern that can happen generally in real-world applications. We have shown the resulting representation is well-suited for both binary and multi-class classification tasks. 
    \item \textit{Transferability:} Different datasets in the intrusion detection task may vary in their input pattern features. We demonstrate that by using masking as an augmentation method, we can handle input features differences among datasets and result in better transferability performance.
    \item \textit{Labeled Dataset Size:} As the main benefit of SSL, the labeled dataset size can be reduced while the performance is maintained to some extent. In this paper we thoroughly investigate labeled dataset size over IDS performance. The results have shown promising performance when the labeled dataset size  is limited.
    \item \textit{Hidden vs Context Representations:} As contrastive learning framework provides two representations for each input pattern, namely hidden and context representations. In this paper we compare the performance in downstream tasks when either hidden or context representation is used.
\end{itemize}

\section{Prior Art}

% \begin{prior_art}%
\noindent
% \section{prior_art}
Basically, an IDS is a network security technology for detecting vulnerability exploits or malicious traffic on a network. Several researches have been done on this topic to distinguish between normal and malicious traffic as a binary classification and also, to detect  attack type as a multi-class classification problem.

\subsection{Supervised Learning} \label{sssec:supervised}
In supervised learning, the main strategy is to learn an input-output mapping based on the (input, output) training pairs \cite{ahmad2021network}. kNN is one of the simplest supervised learning algorithms that assigns the label for each test sample based on majority voting of $k$ nearest training samples to the test sample in the input feature space {\cite{ahmad2021network}}. Kasongo \etal~{\cite{kasongo2020deep}} design IDS based on kNN while they proposed a new wrapper-based feature extraction unit and tested it on UNSW-NB15 {\cite{moustafa2015unsw}} and the AWID \cite{kolias2015intrusion} datasets.

SVM is a supervised ML algorithm based on the idea of max-margin separating hyper-plane in n-dimensional feature space for both linear and nonlinear problems \cite{ahmad2021network}. Jing \etal, \cite{jing2019svm} used a new nonlinear scaling method which is independent of data values as a pre-processing technique on UNSW-NB15 dataset to increase accuracy of SVM-based classifier.

DT is one of the basic supervised ML algorithms which automatically selects the best features for building a tree and then prune it to avoid the over-fitting \cite{ahmad2021network}. Kasongo \etal, \cite{kasongo2020performance} used XGBoost feature selection algorithm to reduce the feature space from $42$ to $19$ on UNSW-NB15 dataset and built DT based on selected features and show the effectiveness of this selection.

RF is one of the mostly used supervised learning algorithms  suitable for classification and regression tasks. It contains trees built by DT algorithm \cite{faker2019intrusion}. Ahmad \etal~\cite{ahmad2021intrusion} proposed feature clusters in terms of Flow, Message Queuing Telemetry Transport (MQTT) and Transmission Control Protocol (TCP) by using features in UNSW-NB15 dataset. Then they used top contributing features selected from TCP, Flow and  MQTT features set with different supervised learning classifiers including RF to increase accuracy.

Ensemble Methods (EM) combines different classifiers with various strengths and weaknesses and get better performance by combining the predictions from multiple models \cite{ahmad2021network}. Rashid \etal~\cite{rashid2022tree} introduced a tree-based Stacking Ensemble Technique (SET) and test the model on NSL-KDD \cite{ibrahim2013comparison} and UNSW-NB15 datasets. They used EM to handle issue of using single classifier on large scale datasets.

ANN is also a supervised ML algorithm and is inspired by the biological neural networks of human brains. ANN can construct nonlinear modeling by learning from larger datasets \cite{ahmad2021network}. Saber \etal~\cite{saber2017performance} provide an optimized ANN for pattern recognition to detect different attacks of KDD CUP99 for multiclass classification problem. To select the important parameters, they have given some and all of the basic attributes to the networks to verify the dependence between the parameters and attack types. Then, the parameters relating to content and time-based ones have been added to demonstrate their utility and performance.

The major commonality in the aforementioned methods is their shallow architecture, which avoids them utilizing dataset information when its size increases, while deep architectures have shown superior performance versus dataset size \cite{dlgb16}. In the sequel, we survey deep architectures.

DNN is a basic DL structure which is used to model complex nonlinear functions by learning in hierarchical layers \cite{ahmad2021network}. Ahmed \etal~\cite{ahmed2020deep} worked on collective anomaly detection problem on UNSW-NB15 and  KDD CUP99 \cite{tavallaee2009detailed} datasets. Due to high FAR in unsupervised methods, they used DL methods which is supervised in nature and other classical learning algorithms such as DT, Auto MLP (AM) and NB for comparison. They have shown that DL outperforms a wide range of unsupervised techniques.

RNN extends the capabilities of the traditional feed-forward neural network and is designed to model the sequence data. For IDS, RNN can be used for the supervised classification and feature extraction. LSTM is a variation of RNN that handle short-term memory problem in long-term sequences \cite{ahmad2021network}. Gwon \etal~\cite{gwon2019network} proposed models based on sequential information using LSTM network and categorical information using  embedding technique.

Convolutional Neural Network (CNN) is another DL structure which consists of an input layer, the stack of convolutional and pooling layers for feature extraction, and finally fully connected layer(s) for classification tasks. CNNs have been used in IDS for the supervised feature extraction and classification purposes \cite{ahmad2021network}. Mulyanto \etal~\cite{mulyanto2020effectiveness} proposed a cost-sensitive neural network based on focal loss, called the Focal Loss Network Intrusion Detection System (FL-NIDS), to overcome the imbalanced data problem. FL-NIDS was applied using DNN and CNN and evaluated on three imbalanced datasets: NSL-KDD, UNSW-NB15, and Bot-IoT \cite{koroniotis2019towards}. \cite{kolosnjaji2016deep}.

In deep learning, a large amount of data must be used for training the model to reach high accuracy. In supervised learning, all data must be labeled, and the lack of labeled data or the low accuracy of the labels reduces the accuracy which necessitates the unsupervised and self-supervised learning strategies.

\subsection{Unsupervised Learning} \label{sssec:unsupervised}

Stacked AutoEncoder (SAE) is a variation of AE which is an unsupervised learning algorithm \cite{ahmad2021network}. Khan \etal, \cite{khan2019novel} proposed a novel two-stage deep learning (TSDL) model, based on a SAE on KDD CUP99 and UNSW-NB15 datasets. The model had two decision stages: the first stage classified network traffic as normal or abnormal. This result is given to the second stage as an additional feature to predict the normal state and other classes of attacks as the final result. The proposed model is able to learn useful feature representations from large amounts of unlabeled data and classifies them automatically and efficiently.

Variational AutoEncoder (VAE) is a variation of AE which is used to model the generative distribution of a set of unlabeled samples \cite{ahmad2021network}. Yang \etal~\cite{yang2019improving} proposed improved conditional VAE (ICVAE) with a DNN, namely ICVAE-DNN that evaluates on NSL-KDD and UNSW-NB15 datasets. ICVAE is used to learn and explore potential sparse representations between network data features and classes. The trained ICVAE decoder generates new attack samples according to the specified intrusion categories to balance the training data and increase the diversity of training samples to improve the DR of the imbalanced attacks. The trained ICVAE encoder is used to automatically reduce data dimension and initialize the weight of DNN hidden layers to improve the fine tuning process. They also use some oversampling methods like random over sampler (ROS), SMOTE, and ADASYN.

Deep Belief Network (DBN) is a DL model constructed by stacking many RBMs followed by a softmax classification layer. This network is mostly used for feature
extraction and classification tasks in IDS. DBN is pretrained using the greedy layer-wise learning approach in an unsupervised manner, followed by a supervised fine-tuning methodology for learning useful features \cite{ahmad2021network}. Tian \etal~\cite{tian2020intrusion} proposed a model based on improved DBN to solve overfitting, low classification accuracy, and high FPR problems.

\subsection{Self-supervised Learning} \label{sssec:self-supervised}
In self-supervised learning the labels are not used during training (similar to unsupervised learning and different from supervised version) while the targeted task is a classification one (similar to supervised learning and different from unsupervised version). 

Wang \etal, \cite{wang2021network} proposed label-free self-supervised learning-based approaches called BYOL. They also used a new data augmentation strategy to learn invariant feature representation capability. The BYOL model is trained on the UNSW-NB15 dataset in a self-supervised manner and network traffic feature representations are extracted. Then the resulting feature extractor model is tested on NSK-KDD, KDD CUP99, CIC IDS2017 \cite{sharafaldin2018toward}, and CIDDS\_001 \cite{ring2017flow} datasets by training a head over the feature extractor separately.

While self-supervised learning is introduced for the first time for IDS in \cite{wang2021network}, the augmentation used is not well-justified. Using a pre-processing method, the authors have argued that they can convert network traffic vector to grayscale image and apply some famous image augmentations such as horizontal flip, vertical flip, random crop and random shuffle to generate augmented data. They claimed that the network traffic after data augmentation retains the original traffic characteristics while data augmentation  introduces different disturbances leading  to train a general model. But this hypothesis is not tested in the paper. 

In all datasets with various features types, missing data can be assumed. Thus, applying masking as a data augmentation method is well justified and can be used to generate augmented data. We use this augmentation in our proposed system. As we have some non-overlapping features in different datasets, using masking can model these features when transferability is desired. We also surveyed the effect of limited labeled data in the accuracy of IDSs when self-supervised learning is adopted for feature extraction.

\section{Proposed method}

\begin{figure*}
  \includegraphics[width=\textwidth,height=5cm]{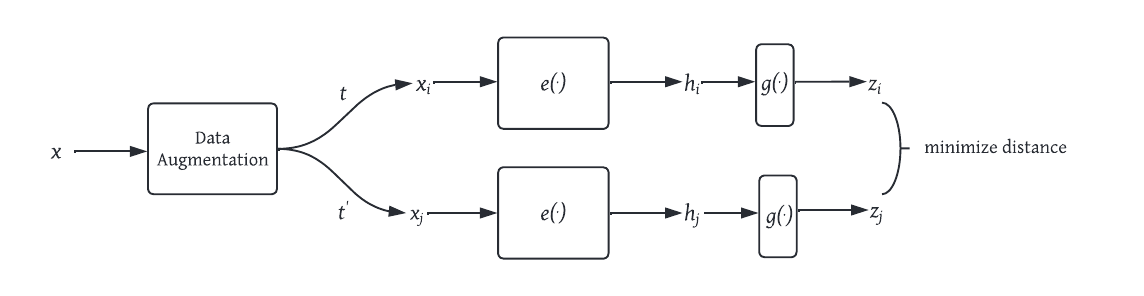}
  \caption{Overview of the model.}\label{fig:SSL}
\end{figure*}

Our proposed method for IDS is based on self-supervised contrastive learning (SSCL). This section presents a detailed description of this method. SSCL trains a model to generate informative representation of input patterns using unlabeled dataset. The prominent goal of SSCL is learning transferable knowledge from unlabeled data and then applying the learned knowledge for downstream tasks \cite{wu2021self}. Figure \ref{fig:SSL} represents a general SSCL architecture. As we can see, the input pattern $\xb$ is fed into the Data Augmentation block where two augmented versions $\xb_i$ and $\xb_j$ are generated. The augmented versions are new viewpoints to the original input pattern and are similar in nature. Then $\xb_i$ passes through $\mathrm{e}(\cdot)$ (encoder block) and $\mathrm{g}(\cdot)$ which generates $\hb_i$ (hidden) and $\zb_i$ (latent) representations, respectively. $\xb_j$ also passes through the same blocks which generates $\hb_j$ and $\zb_j$ representations. Finally $\mathrm{e}(\cdot)$ and $\mathrm{g}(\cdot)$ are trained to minimize the distance between the representations $\zb_i$ and $\zb_j$ of the augmented versions.

\subsection{Data Augmentation}

In this paper, we propose to use masking for data augmentation  where different views of the same input pattern are attained by randomly masking a predefined percentage of input pattern features. The data augmentation module transforms any data sample $\xb$ into $\xb_i$ and $\xb_j$ which are two different views of the data. As we mentioned, data augmentation plays a key role in SSCL and the model must train properly to generate similar representations for similar data.

When a typical sample $\xb$ enters \textit{Data Augmentation} module, a predefined percentage ($m$) of its elements are randomly selected and set to zero (masking). This process is done independently to generate augmented samples $\xb_i$ and $\xb_j$.   
Then these pairs are used for training the encoder network with contrastive loss.
\subsection{Contrastive Learning}
The goal of contrastive learning is to learn representations by contrasting positive pairs (augmented versions of similar samples) against negative pairs (different samples). As we can see in Figure \ref{fig:SSL}, the Encoder $\mathrm{e}(\cdot)$ must learn to produce shared representations from the two augmented views of the same sample while contrasting dissimilar samples. Finally the \textit{Encoder} output is fed into $\mathrm{g}(\cdot)$ block, which is generally a linear projection, to generate latent representation. 

In order to train, we first augment each batch of the unlabeled data with a size of $N$ using the \textit{Data Augmentation} module. This results in $2N$ samples. Consider $\xb_i$ and $\xb_j$, the augmented views of one data sample in the input batch, as positive pairs. The contrastive loss between the latent representations corresponding to this positive pair is defined as \cite{chen2020simple}:
\begin{align}
    l_{i,j} = -\log\frac{\exp(sim({\zb_i},{\zb_j})/\tau)}{\sum_{k=1}^{2N}1_{k\not=i}\exp(sim({\zb_i},{\zb_k})/\tau)},
\label{formula1}
\end{align}
where $sim(\cdot,\cdot)$ is cosine similarity metric between input vectors, and $\tau$ is temperature parameter. Minimizing this loss leads to increasing the similarity of latent representation of augmented versions for an input pattern and decreasing similarity of latent representation for augmented versions of different input patterns. Since the batch size is $N$, there are $N$ positive pairs in the output of \textit{Data Augmentation} block. The final loss is formed by considering the distance for all positive pairs as:
\begin{align}
   Loss = \frac{1}{2N}\sum_{k=1}^{N}[ l(2k-1, 2k) + l(2k, 2k-1) ],
\label{formula2}
\end{align}
where $(2k,2k-1)$ refers to the indices of a positive pair and the loss becomes symmetrical by considering both $l(2k-1, 2k)$ and $l(2k, 2k-1)$ terms. The model is trained to minimize the loss until it converges \cite{chen2020simple}.

\begin{algorithm}
\caption{Self-Supervised Contrastive Learning \cite{chen2020simple}}\label{alg:cap}
\begin{algorithmic}[1]
\State \textbf{input:} batch size N, constant $\tau$, masking ratio $m$ and structure of e and g
       \For {sampled minibatch $\{x_k\}_{k=1}^{N}$}
       \For{all $k \in \{1,...,N$ \}}
       \State Draw two augmentation function $t$ and $t'$
       \newline
       \color{gray}
       \hspace{1.3cm} \# The first augmentation
       \color{black}
       \State $\xb_{2k-1}$ = $t$($\xb_k$, $m$)
       \State $\hb_{2k-1}$ = $e$($\xb_{2k-1}$)
       \State $\zb_{2k-1}$ = $g$($\hb_{2k-1}$)
       \newline
       \color{gray}
       \hspace{1.3cm} \# The second augmentation
       \color{black}
       \State $\xb_{2k}$ = $t'$($\xb_k$, $m$)
       \State $\hb_{2k}$ = $e$($\xb_{2k}$)
       \State $\zb_{2k}$ = $g$($\hb_{2k}$)
       \EndFor
       \newline
       \color{gray}
        \hspace{0.6cm} \# calculate contrastive loss
       \color{black}
       \For {all $i$ $\in$ $\{1,...,2N\}$ and $j$ $\in$ $\{1,...,2N\}$} 
       \newline
       \color{gray}
       \hspace{1.3cm} \# pairwise similarity
       \color{black}
       \State {$s_{i,j}$ = $\zb_i{}^{T}\cdot \zb_j/(\parallel \zb_i\parallel \parallel \zb_j \parallel)$}
       \EndFor
       \newline
       \color{gray}
       \hspace{0.6cm} \# define $l(i,j)$
       \color{black}
       \State  $l(i,j) = -\log\frac{\exp(s({\zb_i},{\zb_j})/\tau)}{\sum_{k=1}^{2N}1_{k\not=i}\exp(s({\zb_i},{\zb_j})/\tau)}$
       \newline
       \State$Loss$ = $\frac{1}{2N}\sum_{k=1}^{N}[l(2k-1, 2k) + l(2k, 2k-1)]$
       \State update ${e(\cdot)}$ and ${g(\cdot)}$ to minimize $Loss$
       \EndFor
       \State \textbf{return} ${e(\cdot)}$ and ${g(\cdot)}$
\end{algorithmic}
\end{algorithm}

\subsection{Applications}
Architectures based on SSCL can be used in different applications. The hidden and latent representations ($\hb$ and $\zb$) are the result of training based on unlabeled data and can utilize huge amounts of data. This leads to efficient abstract representations suitable for downstream tasks. In this paper we evaluate the efficiency of these representations for both binary and multi-class classification tasks.

Using an architecture trained with SSCL, we can use the resulting representation for supervised training over small size dataset. This feature is especially useful when collecting labeled data is cost or time consuming. 

As we use masking for data augmentation, the resulting model is robust against missing data in the input pattern. From another point of view, the resulting architecture can be used to extract efficient representations for input patterns coming from other datasets where the features are not exactly similar.

In the next section, we evaluate the performance of the proposed method in different scenarios. 

\section{Simulations}
In this section, we check the performance of the proposed method for IDS. 

\textbf{Dataset and Default setting.} We use the UNSW-NB15 dataset to train our model. As it is shown in Table \ref{tab:DINFO}, two different packs of the data are used in this work, namely Larger Pack and Smaller Pack. The training set is used for training the encoder and we call it the "Encoder set" and the testing set is used for training the classification head and we call it the "Head set". Since this dataset contains rich feature information, it is a suitable choice for training and also for transferring  on other datasets. The Batch size is set to 32 and we use $\tau$ = 0.5 for contrastive loss. We also use AdamW optimizer and exponentialLR scheduler  with learning rate = 0.0002 for training. Two encoder structures are designed for the two packs of data, as shown in Table \ref{tab:encinfo}. The $\mathrm{g}(\cdot)$ block is a linear layer that maps hidden representations to context representations and is shown in Table \ref{tab:encinfo}.

\begin{table} [!htb]
	\caption{UNSW-NB15 dataset}
	\renewcommand{\arraystretch}{1.5}
	%if using array.sty, it might be a good idea to tweak the value of
	%\extrarowheight as needed to properly center the text within the cells
	%\caption{Accuracy over clean test set, number of iterations to converge, training time for a typical sample, maximum %and minimum spectral norm over convolutional and fully connected layers for CNN Architecture over different datasets}
	\label{tab:DINFO}
	\centering
	% Some packages, such as MDW tools, offer better commands for making tables
	% than the plain LaTeX2e tabular which is used here.
	\resizebox{0.48\textwidth}{!}{%
		\setlength\tabcolsep{2pt}
		\begin{tabular}{ccccc}
			\hline
			 &Smaller pack &&Larger pack\\ \hline
			Attack category&Training&Testing &Training&Testing \\\hline
			Normal & 56000 & 37000& 542254& 135531 \\
			Fuzzers & 18184 & 6062 & 4034 &1017\\
			 Analysis& 2000& 677 & 404& 122 \\
			 Backdoors& 1746&583& 426& 108 \\
			 Dos& 12264&4089& 921& 246 \\
			 Exploits& 33393&11132& 4337 & 1072 \\
			 Generic& 40000&18871& 6029 & 1493 \\
			 Reconnaissance& 10491&3496& 1398 & 361 \\
			 Shellcode& 1133&378& 177 & 46 \\
			 Worms& 130&44& 20 & 4 \\
			 \hline
			Total& 175341& 82332& 560000 & 140000\\
			\hline
		\end{tabular}
	}
\end{table}

\begin{table} [!htb]
	\caption{The structure of Encoder($\mathrm{e}(\cdot)$) and $\mathrm{g}(\cdot)$ blocks}\label{tab:encinfo}
	\renewcommand{\arraystretch}{1.5}
	%if using array.sty, it might be a good idea to tweak the value of
	%\extrarowheight as needed to properly center the text within the cells
	%\caption{Accuracy over clean test set, number of iterations to converge, training time for a typical sample, maximum %and minimum spectral norm over convolutional and fully connected layers for CNN Architecture over different datasets}
	\label{tab:reg}
	\centering
	% Some packages, such as MDW tools, offer better commands for making tables
	% than the plain LaTeX2e tabular which is used here.
	\resizebox{0.48\textwidth}{!}{%
		\setlength\tabcolsep{3pt}
		\begin{tabular}{cccccccccc}
		     \hline
			Block&Layers&For Smaller pack&For Larger pack\\\hline
			Encoder ($\mathrm{e}(\cdot)$) &Conv & 32@(1×2) & 8@(1×2)\\
			&Conv & 64@(1×2) & 16@(1×2)\\
			&Conv& 128@(1×2)& 32@(1×2)\\
			&Max Pool& 1×3 &\_\_\\
			&Conv& 256@(1×2)&64@(1×2) \\
			&Max Pool& 1×2 &1×3 \\
			&Conv& 512@(1×2)&128@(1×2) \\
			&Max Pool& 1×4 &1×4 \\
			&Conv& \_\_ &256@(1×2) \\
			\hline
			$\mathrm{g}(\cdot)$&Linear& 512×256 &256×128 \\
			 \hline
			Parameters && 482528& 121456\\
			\hline
		\end{tabular}
	}
\end{table}

\textbf{Preprocessing.} As different datasets for IDS contain categorical features, we encode these features into numerical ones via one-hot encoding. Then MinMax scaler is used for normalizing the data between 0 and 1 as:
\begin{align}
    \xb_{scaled} = \frac{\xb - \xb_{min}}{\xb_{max} - \xb_{min}}.
\label{formula3}
\end{align}
where $\xb$ and $\xb_{scaled}$ represents feature before and after normalization and $\xb_{min}$ and $\xb_{max}$ represents minimum and maximum value of feature across training patterns before normalization. Some data samples are filled with "-" in the service column in our dataset. We mask this value after one-hot encoding.

\textbf{Training Encoder and $\mathrm{g}(\cdot)$ Blocks.} To train these blocks, we use SSCL based on the loss function defined in \eqref{formula2}. The trained blocks are then frozen and their representations are used for downstream tasks.

\textbf{Training Classification Head.} The classification head is simply a fully-connected layer followed by a softmax classification layer. The number of output neurons is selected based on the downstream task.

\subsection{Comparison to state-of-the-art methods}
To compare with the other methods, we use the Smaller Pack for our simulation (Other methods are generally evaluated over this pack). In this experiment, $80\%$ of the unlabeled Encoder set is used for self-supervised training and the rest is used for evaluation. $80\%$ of the Head set is used for training the classifier head. We use the rest of the Head set for evaluation and the results are shown in Table \ref{tab:sota}. For better comparison, the table is divided into two parts where the upper part shows supervised methods and lower part comprises self-supervised methods. In most cases, our model achieves better performance than the other self-supervised method and we have about 4\% and 2\% improvement in accuracy and  F1 score respectively. As we can see, the proposed method outperforms the self-supervised approach suggested in {\cite{wang2021network}}. The main reason is the masking method we suggest for data augmentation in the self-supervised learning part. A different augmented version of the same data must be a valid instance. Masking can be considered a case where some features of the input pattern are missed and thus the resulting masked pattern is valid. On the other hand, using different types of data augmentation generally used for image augmentation can deteriorate the self-supervised training as the augmented version may not be valid. Altogether, in our proposed method, the augmented samples are guaranteed to be valid, leading to better performance for the proposed method.

\begin{table} [!htb]
	\caption{Comparison of different supervised and self-supervised IDSs over UNSW-NB15 dataset}\label{tab:sota}
	\renewcommand{\arraystretch}{1.5}
	%if using array.sty, it might be a good idea to tweak the value of
	%\extrarowheight as needed to properly center the text within the cells
	%\caption{Accuracy over clean test set, number of iterations to converge, training time for a typical sample, maximum %and minimum spectral norm over convolutional and fully connected layers for CNN Architecture over different datasets}
	\label{tab:reg}
	\centering
	% Some packages, such as MDW tools, offer better commands for making tables
	% than the plain LaTeX2e tabular which is used here.
	\resizebox{0.48\textwidth}{!}{%
		\setlength\tabcolsep{2pt}
		\begin{tabular}{cccccccccc}
			\hline
			Method & Name & Accuracy & Precision & Recall& F1 score\\ \hline
			Supervised & VLSTM \cite{zhou2020variational} &\_\_&0.8600&\textbf{0.9780}& 0.9070 \\
			 & MFFSEM \cite{zhang2021multi} &0.8885&0.9388&0.8044& 0.8664 \\
			 & TSIDS \cite{tian2021two} & \_\_&0.9516&0.9515 & 0.9515 \\
			 & SADE-ELM \cite{wang2021intrusion}& 0.7238 &0.6994&0.8742& 0.7771 \\
			 & BoTNet \cite{wang2021network}& \textbf{0.9405} &\textbf{0.9618}&0.9447& \textbf{0.9532} \\
			 \hline
			Self-supervised & BoTNet \cite{wang2021network} & 0.8997 &0.8972&\textbf{0.9526}& 0.9241\\
			 & Our model & \textbf{0.9419} &\textbf{0.9451}&0.9419& \textbf{0.9418}\\
			\hline
		\end{tabular}
	}
\end{table}

\subsection{Limited labeled data}
One of the most important benefits of SSCL is using huge amounts of easily accessible unlabeled data for training and producing informative representations and then use it for the classification of  small size labeled datasets.
In this experiment, we check this feature over the Larger Pack. For training the classification head, we use a different ratio of the Head data for each attack class. For binary classification, there are 2 classes and in multi-class classification, we use 6 classes which are "Normal", "Fuzzers", "Dos", "Exploit", "Generic" and "Reconnaissance". The results are illustrated in Table \ref{tab:ratio}. As we can see, our model performance is almost retained when the size of the labeled dataset is decreasing (Note that in all cases, the Encoder Block is similar). In the extreme case with just 1\% of the labeled data, the accuracy decreases about 0.12\% in comparison to the case where all the labeled dataset is used for classification head in binary classification.

\begin{table} [!htb]
	\caption{Comparison of different labeled dataset size over the performance of IDSs}\label{tab:ratio}
	\renewcommand{\arraystretch}{1.5}
	%if using array.sty, it might be a good idea to tweak the value of
	%\extrarowheight as needed to properly center the text within the cells
	%\caption{Accuracy over clean test set, number of iterations to converge, training time for a typical sample, maximum %and minimum spectral norm over convolutional and fully connected layers for CNN Architecture over different datasets}
	\label{tab:reg}
	\centering
	% Some packages, such as MDW tools, offer better commands for making tables
	% than the plain LaTeX2e tabular which is used here.
	\resizebox{0.48\textwidth}{!}{%
		\setlength\tabcolsep{2pt}
		\begin{tabular}{cccccccccc}
			\hline
			Classification & data ratio & Accuracy & Precision & Recall& F1 score  \\ \hline
			Binary& 100\%&0.9943&0.9965&0.9943&0.9950\\
			& 10\% &0.9941 &0.9970&0.9941& 0.9952\\
			  & 5\% &0.9941&0.9969&0.9941 &0.9952\\
			 & $1\% $ & 0.9931&0.9958&0.9931&0.9939 \\
			\hline
			6 classes& 100\%&0.9886&0.9919&0.9886&0.9900\\ & 10\% & 0.9869 &0.9902&0.9869& 0.9882 \\
			& 5\% &0.9861&0.9882&0.9861 &0.9868\\
			 & $1\% $ & 0.9779&0.9833&0.9778 &0.9801 \\
			\hline
		\end{tabular}
	}
\end{table}

\subsection{Transferability of Representations}
As we mentioned, when the Encoder block is trained over an original dataset ($\mathcal{D}_O$), it can be used to generate abstract representations over other datasets ($\mathcal{D}_T$). For this purpose, we need to omit the features available in set $\Dc_T-\Dc_O$  and mask the features in set $\Dc_O-\Dc_T$ ($\Dc_T-\Dc_O$ represent the set of features in transfer dataset that are not in the original dataset). In this experiment, we evaluate Encoder Block representation transferability across datasets. For this experiment, the Larger Pack is used. We consider three datasets for transfer as:

\begin{itemize}
    \item \textit{CIC IDS2017 dataset}: A traffic network dataset with 692703 records includes one normal class and five attack classes. We use 80\% of the data for training the classification head and 20\% for evaluation.
    
    \item \textit{CIDDS 001 dataset:} This dataset consists of traffic data from two External and OpenStack servers. We use data captured by the External server which consists of 671241 records where 648000 records were used to train the classifier head and 23241 records were used for evaluation.
    
    \item \textit{BoT-IoT dataset:} The Bot-IoT dataset \cite{koroniotis2019towards} was collected using smart home appliances in a lab environment. Traffic samples collected for Industrial IoT experiments are also included in this dataset. Temperature monitoring systems, freezers, kitchen appliances, and motion-controlled lights are among the smart home appliances. We used approximately 3.6 million records from the full Bot-IoT dataset. In five percent of the dataset, the top ten features extracted from the raw data are organized into five main classes: DDoS, Dos, Reconnaissance, Theft, and Normal. The training and test packs contain 2934817 and 733705 samples, respectively.
\end{itemize}

The simulation results are shown in Table \ref{tab:transfer}.
In CIS IDS2017 dataset, our model improves accuracy and F1 score by about 1\% and about 2\%, respectively, in comparison to  BoTNet. We have also found that our model has less than 1\% improvement in accuracy and F1 score over BoTNet's model for the CIDDS 001 dataset. Compared to the standard evaluation criteria, we see an improvement in all evaluation criteria for the BoT-IoT dataset. This experiment represents high transferability of Encoder Block representation which originates from the masking operation used for augmentation. This operation can be efficient to handle the variations across datasets features. 

\begin{table} [!htb]
	\caption{Comparison of transferability of designed system to CIC IDS2017, CIDDS 001 and BoT-IoT datasets}\label{tab:transfer}
	\renewcommand{\arraystretch}{1.5}
	%if using array.sty, it might be a good idea to tweak the value of
	%\extrarowheight as needed to properly center the text within the cells
	%\caption{Accuracy over clean test set, number of iterations to converge, training time for a typical sample, maximum %and minimum spectral norm over convolutional and fully connected layers for CNN Architecture over different datasets}
	\label{tab:reg}
	\centering
	% Some packages, such as MDW tools, offer better commands for making tables
	% than the plain LaTeX2e tabular which is used here.
	\resizebox{0.48\textwidth}{!}{%
		\setlength\tabcolsep{3pt}
		\begin{tabular}{cccccccccc}
			\hline
			Name & method & Accuracy& Precision & Recall& F1 Score \\ \hline
			CIC IDS2017 & BoTNet \cite{wang2021network} &0.9670&0.9500&0.9596&0.9548\\
			& Our model &\textbf{0.9775}&\textbf{0.9791}&\textbf{0.9775}&\textbf{0.9775}\\
			\hline
			CIDDS 001 & BoTNet \cite{wang2021network} &0.9813&0.9816&\textbf{0.9953}&0.9884 \\
			&Our model &\textbf{0.9929}&\textbf{0.9934}&0.9929&\textbf{0.9924}\\
			\hline
			BoT-IoT & TSODE \cite{fatani2021iot} &0.9904&0.9904&0.9904&0.9904 \\
			&Our model &\textbf{0.9983}&\textbf{0.9983}&\textbf{0.9983}&\textbf{0.9982}\\
			\hline			
			
		\end{tabular}
	}
\end{table}

\subsection{Hidden vs Context representations}
As we can see in Figure \ref{fig:SSL}, two hidden and context representations are generated for each input pattern. In this experiment we compare the result of downstream tasks when either hidden or context representations are used. The results on Smaller Pack are shown in Table \ref{tab:hvc}. As we can see, in both tasks and all metrics, choosing hidden representation leads to better performance.

\begin{table} [!htb]
	\caption{Comparison of IDS performance while  using hidden or context representations}\label{tab:hvc}
	\renewcommand{\arraystretch}{1.5}
	%if using array.sty, it might be a good idea to tweak the value of
	%\extrarowheight as needed to properly center the text within the cells
	%\caption{Accuracy over clean test set, number of iterations to converge, training time for a typical sample, maximum %and minimum spectral norm over convolutional and fully connected layers for CNN Architecture over different datasets}
	\label{tab:reg}
	\centering
	% Some packages, such as MDW tools, offer better commands for making tables
	% than the plain LaTeX2e tabular which is used here.
	\resizebox{0.48\textwidth}{!}{%
		\setlength\tabcolsep{2pt}
		\begin{tabular}{cccccccccc}
			\hline
			Classification& Representations & Accuracy & Precision & Recall& F1 score \\ \hline
			Binary &Hidden & \textbf{0.9491} &\textbf{0.9527}&\textbf{0.9491}&\textbf{0.9492}\\
			&Context& 0.9419&0.9451&0.9419&0.9418\\
			\hline
			6 classes& Hidden&\textbf{ 0.8972} &\textbf{0.9108}&\textbf{0.8973}&\textbf{0.8969}\\
			&Context& 0.8890&0.9054&0.8890&0.8890\\
			\hline
		\end{tabular}
	}
\end{table}
\section{Conclusions}
In this paper, we propose a method for intrusion detection based on deep neural networks. In the proposed method, we trained an Encoder Block based on self-supervised contrastive learning using unlabeled training patterns. The resulting representation is then fed into a classification head which is trained using a labeled dataset. The proposed method has shown state of the art results in both binary and multi-class classification tasks among the methods that use self-supervised learning. We have also demonstrated the efficiency of the proposed method when the number of labeled training patterns is limited. The proposed method presents competitive results when only $1\%$ of labeled training patterns are used. Finally we have shown that the trained Encoder Block can be used to produce efficient hidden representations for input patterns coming from datasets different from the dataset used for Encoder Block training. Due to transferability and the ability to handle small size labeled dataset, the proposed method can be efficiently used to detect new class of intrusion using limited labeled samples over different datasets. This feature makes it an interesting option for real-world applications.

	\ifCLASSOPTIONcaptionsoff
	\newpage
	\fi

	
	
	
	\bibliographystyle{IEEEtran}
	\bibliography{myref}
\end{document}